\documentclass[11pt,oneside,english,journal=mamobx,manuscript=article,maxauthors=15,biblabel=plain]{amsart}
\usepackage[T1]{fontenc}
\usepackage[latin9]{inputenc}
\usepackage[letterpaper]{geometry}
\usepackage{amsthm}
\usepackage{graphicx}
\usepackage{setspace}
\makeatletter
\numberwithin{equation}{section}
\numberwithin{figure}{section}

\makeatother

\usepackage{babel}
\begin{document}

\title{Short Cyclic Structures in Polymer Model Networks: A Test of Mean
Field Approximation by Monte Carlo Simulations}

\author{M. Lang, K. Schwenke, and J.-U. Sommer}
\begin{abstract}
A mean field rate theory description of the homo- and co-polymerization
of $f$-functional molecules is developed, which contains the formation
of short cyclic structures inside the network. The predictions of
this model are compared with Monte-Carlo simulations of cross-linking
of star polymers in solution. We find that homo-polymerizations are
well captured by mean-field models at concentrations larger than one
quarter of the geometrical overlap concentration. All simulation data
can be fit using a single geometric parameter for cyclization. The
simulation data reveal that within the range of parameters of the
present study correlations among multiply connected molecules can
be neglected. Thus, mean-field treatments of homopolymerizations are
reasonable approximations, if short cycles are properly addressed.
Co-polymerization is considered in the case of strict A-B reactions,
where all reactive groups of individual molecules are either of type
A or B. For these systems we find a clear influence of the local intermixing
of A and B groups for all concentrations investigated. In consequence,
mean-field models are less appropriate to describe the simulation
data. The lack of ring structures containing an odd number of molecules
as compared to homopolymerizations at same extent of reaction allows
for the formation of stable AB networks at concentrations one order
of magnitude below the geometrical overlap concentration.
\end{abstract}
\maketitle

\section{Introduction}

Cyclization during stepwise polymerization is a still incompletely
understood problem. While there are clear experimental data and well
accepted theoretical models for systems resulting in linear chains
and rings, the situation is much more complicated for network forming
reactions \cite{Sarmoria}. There, only indirect experimental measures
for the amount of ring formation inside the gel are available: for
instance, the weight distribution of sol or the shift in the gel point
can be used to estimate the degree of cyclization inside the gel \cite{Suematsu}.
Both methods require a theoretical model based upon several assumptions
to extract the degree of cyclization from the available data. Therefore,
a rigorous quantitative description of the processes of branched polycondensation
with allowance of intramolecular reactions is not yet available \cite{Kuchanov}.

Computer simulations, on the other hand, allow to directly analyze
the degree of cyclization for any cross-linking reaction. Therefore,
a large number of simulation works in the past focused on the effect
of cyclization as summarized in \cite{Suematsu,lang2005intramolecular}.
While dynamic simulations of network formation mainly concentrate
on the total amount of cycles or the shortest cycles in the sample,
the data focusing on critical dilution was obtained either by experiments
or static simulations of a percolating lattice. A simulation study
that includes the dynamics of the molecules and thus, incorporates
the effects of flocculation at concentrations towards critical dilution,
however, is still lacking.

In general, any theoretical description of network formation is founded
on several assumptions concerning the reaction mechanism, the motion
of polymers, mixing of different species, etc ..., which are difficult
to test by experiments \cite{Kuchanov}. Since computer simulations
can model these effects and allow for a detailed analysis, a critical
test of the assumptions underlying the theoretical models is of great
importance, but still remains incomplete in previous works \cite{lang2005intramolecular,LangSommer08,LaySommerBlumen99,trautenberg1995structure}.

As reviewed in \cite{Kuchanov}, there are several approaches for
a theoretical treatment of the problem of cyclization. Spanning tree
models \cite{Sarmoria,Gordon,Sarmoria2,Dusek,Dusek2} essentially
treat ring forming reactions by assigning pairs of unreacted sites
inside the branched molecules. This assignment is typically based
upon estimating the return probability of Gaussian chains or branched
structures with Gaussian conformations. Conformational changes upon
ring formation and correlations among rings are usually neglected
and the total effect of ring formation is expressed in a shift of
the total conversion. Beyond the gel point, this general scheme is
no more applicable, since the probability for ring formation is equal
to one within this approach, while the actual rate of intramolecular
reactions is rather proportional to the square number fraction of
reactive groups inside the gel. Sarmoria et al. \cite{Sarmoria2}
therefore suggest three different models that circumvent this problem,
but due to the lack of suitable experimental or simulation data, it
is difficult to decide, which model might be the best approximation
for real systems.

Rate theories \cite{Stanford1,Stanford2} use a set of differential
equations that describe the transition among different species in
the reaction container. The advantage of this method is that it is
exact within a mean-field approach and that it allows to model any
correlation among the particular molecular species or conformational
changes induced by cyclization. It has been shown \cite{lang2005intramolecular},
that the predictions of rate theories and spanning tree models or
the work of Suematsu \cite{Suematsu} are consistent with simulation
data and among each other at low degrees of cyclization \cite{Kuchanov,lang2005intramolecular}.
But as a mean-field treatment, such methods are not sensitive to fluctuations.
Thus, a cross-test of rate theory (or any other equivalent mean-field
approach) with computer simulations of network formation that explicitly
models mixing, dynamics, and fluctuations would be important for testing
the applicability of mean-field models to the problem of cyclization
inside a gel. This test will be most critical at high rates of loop
formation at or below the overlap concentration. Therefore, we study
in the present work the rates of cyclization as function of concentration
and compare with rate theory predictions in which we incorporate all
necessary information on molecular conformations, concentrations,
mixing, and connectivity of the molecules.

As model systems we discuss the homopolymerization and co-polymerization
of structurally identical $f$-functional molecules. The latter is
the simplest example of reacting molecules that are subject to composition
fluctuations of reactive groups of different type. As further simplification
we investigate star-shaped molecules with arms of equal degree of
polymerization $N_{a}=N/f,$ where $N$ is the degree of polymerization
of the molecule. Recently, T. Sakai \textit{et al.} synthesized hydrogels
made of 4-arm star polymers with tetra-Nhydroxysuccinimide-glutarate-terminated
PEG (A-type) and tetraamine-terminated PEG (B-type) as precursor molecules,
``Tetra-PEG-gels'' \cite{Sakai08}. These gels, which showed remarkable
mechanical properties, should have a structure equivalent to the co-polymer
model networks of our study.

This work is structured as follows. First, we discuss the reaction
rates used to model gelation. Then, we compute the distribution of
species concerning connectivity inside the evolving network. The network
properties are estimated based on a simplified phantom model picture.
All theoretical predictions are compared with the simulation data
obtained by cross-linking simulations of solutions of star molecules
as described in the previous paper \cite{Schwenke}. Note that these
simulations explicitly model dynamics and conformations of all molecules
or clusters of molecules during cross-linking. Furthermore, excluded
volume, the concentration dependence of chain conformations, and network
elasticity is explicitly modeled, which is a clear advantage over
previous simulation studies on cyclization in networks \cite{Leung,Cail,Stepto}.
The main parameters of the simulation data we have to keep in mind
are that a) all reactions were stopped close to 95\% conversion of
the reactive groups, b) all data is for 4-functional stars and c)
all data is always represented as function of concentration $c$ with
respect to overlap concentration $c^{*}.$ Therefore, all theoretical
computations below use the same conversion $p=0.95$ for comparison,
which is close to previous NMR results \cite{Lange2}. As in previous
works \cite{trautenberg1995structure,lang2005intramolecular,lang2004promotion},
we did not observe significant modifications (except of some weak
deviations for AB networks at $p>0.9$) of network structure as function
of the reactivity of the molecules. Therefore, we compare the data
of the present study that is obtained at reaction probability one
(limit of diffusion-controlled reactions) with theoretical predictions
that are derived for equilibrium reactions at infinitely low reaction
rate (limit of reaction-controlled reactions) without any further
corrections. In order to eliminate time dependencies, all data and
theory is presented as function of conversion. Note that the overlap
concentration $c^{*}$ is determined by a geometrical analysis of
chain conformations as described in ref \cite{Schwenke}.

\section{Reaction rates\label{sec:Reaction-rates}}

\begin{figure}
\begin{center}\includegraphics[width=0.85\columnwidth]{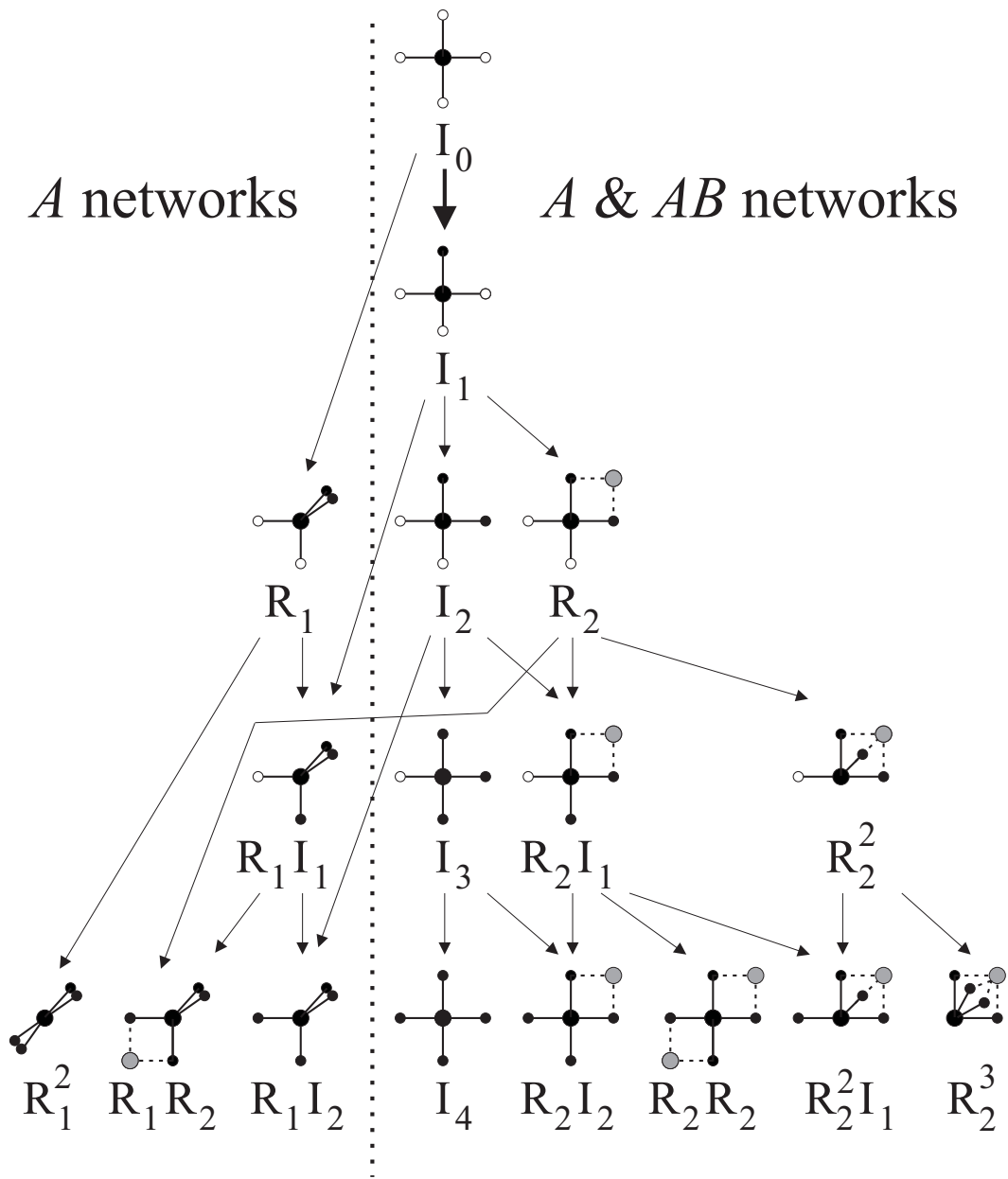}\end{center}

\caption{\label{Flo:Reaction diagram}Reaction diagram: $f$-functional molecules
(stars) are depicted by their centers (lage black circles) and their
ureacted (small white circles) and reacted (small black circles) groups
that can form multiple connections to neighboring stars (grey circles).
Star arms are shown by black lines, arms of neighboring stars by dashed
lines. The arrows indicate the possible irreversible reactions.}
\end{figure}

First, we develop a notation scheme that allows to distinguish any
possible local connectivity of a given $f$-functional molecule. Throughout
this work we restrict the analysis to nearest neighbour connections
inside the gel. Therefore, only cyclic structures containing one or
two molecules are distinguished from all remaining connections. Let
$R_{i}$ denote that a molecule is attached to a ring containing $i$
molecules. Similarly we use $R_{i}^{j}$ to describe that the molecule
is part of $j$ rings of size $i$, whereby we distinguish between
different neighbours for ring formation. Thus, $R_{2}R_{2}\neq R_{2}^{2}$,
since the former describes two ``double links'' $R_{2}$ with two
\emph{different} neighbours, while the latter is a ``triple link''
with the \emph{same} neighbour, see Figure \ref{Flo:Reaction diagram}
for more details. $I_{j}$ describes an $f$-functional star that
forms $j$ ``ideally'' branching connections; not yet connected
groups are not denoted. All available information on the connectivity
of a given star as shown in Figure \ref{Flo:Reaction diagram} is
summarized in the notation $R_{1}R_{2}$, $R_{1}I_{2}$, etc ...,.
Note that throughout this section we consider only number fractions
of species among all stars. Thus, the star connectivity $R_{1}R_{2}$
simultaneously denotes the number fraction of stars with exactly this
connectivity.

Let us also introduce reaction rates $k_{l,m}$ to describe irreversible
transitions among the different possible connectivities of the stars.
Here, the index $l$ is $0,$ if the reaction is ideally branching,
or equal to $i$, if a ring of $i$ molecules is formed. The index
$m=0,...,f-1$ denotes the number of reacted groups at the molecules
for $l=0,1$ and the number of existing bonds between the given pair
of stars for $l=2$, see Figure \ref{Flo:Reaction diagram} and equations
(\ref{eq:di0}) to (\ref{eq:dr1r2}) for more details. 

For deriving the rate constants among the different transitions we
use the following rules:

a) All reaction rates are estimated based upon local concentrations
of the particular pairs of reactive groups.

b) Reactions among previously unconnected groups are independent of
each other.

c) Reactions among previously connected groups depend on the equilibrium
conformations of the structure between the reactive groups.

For the \emph{ideally branching reactions }we assume that the reaction
rate of a given reactive group is proportional to the average concentration
of all reactive groups at unconnected stars, which is approximately
$cf(1-p)$. Here, $cf$ is the initial concentration of reactive groups
and $p$ the fraction of reacted groups. Since in the present work
we only consider connections up to next neighbours, this approximation
is also applicable beyond the gel point \cite{lang2004promotion}.
The reaction rate of a star with $f-m$ unreacted groups is then given
by 
\begin{equation}
k_{0,m}(c)=cf(f-m)(1-p).\label{eq:skl}
\end{equation}
For the above equation (and the equations below) we implicitly assume
that steric repulsion among reactive groups can be neglected, which
is a reasonable approximation for stars of small $f$. Networks of
large $f$ will require an improvement of the above equation.

The \emph{formation of loops $R_{1}$} is computed by estimating the
concentration of pairs of unreacted groups inside a star. For\textbf{
$c>c^{*},$} the star conformation changes as function of concentration
as discussed in our previous work \cite{Schwenke}, and thus, the
rate of collisions between two reactive groups of a star depends on
the size of the star
\begin{equation}
k_{1,m}(c>c^{*})\approx k_{1,m}(c^{*})\cdot(c/c^{*})^{-\beta}.\label{eq:k1m}
\end{equation}
We find $\beta\approx0.65$, if we only consider the concentration
dependence of star conformations. A best fit of the loop data for
$c>c^{*}$ reveals $\beta\approx0.53$, which might be caused by an
additional effect of desinterpenetration of stars, since also the
shape of the correlation hole of the star centers changes as function
of concentration \cite{Schwenke}. For $c<c^{*}$ we assume that the
conformation of the star and thus, the frequency of collisions inside
the star remain unchanged as function of concentration: 
\begin{equation}
k_{1,m}(c<c^{*})\approx k_{1,m}(c^{*}).\label{eq:k1m2}
\end{equation}
The frequency of collisions of reactive groups inside a star at a
given concentration $c$ is proportional to the number of distinguishable
pairs of arms and the single pair reaction rate $k_{1,f-2}(c)$, 
\begin{equation}
k_{1,m}(c)=\left(\begin{array}{c}
f-m\\
2
\end{array}\right)k_{1,f-2}(c)\label{eq:k1n}
\end{equation}
which we denote using a binomial coefficient for the number of pairs
of unreacted arms.

At the geometrical overlap concentration $c=c^{*}$ and at the beginning
of the reaction, $p=0$, we fix the ratio of single pair intra-molecular
reactions $k_{1,f-2}$ and ideal reactions of a single reactive group,
$k_{0,f-1}$, by comparing the concentrations of the reaction partners
\begin{equation}
\frac{k_{1,f-2}(c^{*})}{k_{0,f-1}(c^{*})}\approx\frac{R_{g}^{3}}{R_{e}^{3}}.\label{eq:k1f2}
\end{equation}

\begin{figure}
\includegraphics[width=0.5\columnwidth]{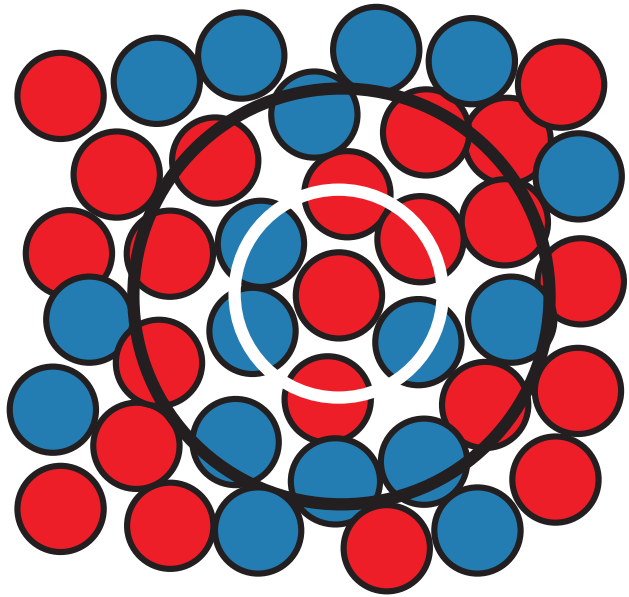}

\caption{\label{fig:The-gyration-volumes}The gyration volumes $R_{g}^{3}$
densely fill space near $c^{*}$. Stars of different type are displayed
in different color. The white circle indicates the volume $R_{e}^{3}$
from which reactive groups are selected for reaction, the black shows
the range within the centers of mass of possible reaction partners
must be located.}
\end{figure}

This relation reflects that the geometrical overlap condition is determined
from changes in $R_{g}$, while reactions can occur within the volume
that is accessible to the ends groups attached to the arms of the
stars, see Figure \ref{fig:The-gyration-volumes}. If we use the ideal
prediction for $R_{g}^{2}=(3-2/f)\cdot N_{a}b^{2}/6$ and estimate
$R_{e}\approx2^{1/2}N_{a}$, we find an additional geometrical prefactor
of approximately 6/5 for the present case of 4-arm stars in order
to fit the simulation data on the formation of loops $R_{1}$. We
note that $R_{g}^{3}/R_{e}^{3}$ as computed by theory or by directly
taking the simulation data differ by less than 2\% for $N\ge16$.
Therefore, the prefactor of approximately 6/5 must be attributed to
the partial disinterpenetration of the stars at $c^{*}$ due to excluded
volume interactions between stars. This is also sketched at Figure
\ref{fig:The-gyration-volumes}: repulsion and disinterpenetration
among stars occurs up to $R_{g}$, while reaction partners are selected
via the endgroups on a larger distance as indicated by the white circle.

The\emph{ reaction rates for loops $R_{2}$} are estimated by computing
the concentrations of \emph{pairs} of reactive groups among two different
connected stars. For any star we know the functionality $f$, the
number of reacted groups $m$, the number of different connected stars
$b\le m$, and the number $n$ of bonds to a particular connected
star. The missing information is the average number of unreacted groups
$b_{n}$ at each connected star, if there are $n$ bonds connecting
to this star for computing the number of unreacted pairs that form
loops $R_{2}$. Here, $b_{n}$ is derived from the reaction diagram
by considering that any star is selected proportional to the weight
fraction of its reacted arms (or pairs of arms, if there are multiple
connections). The expected number of unreacted groups can be written
as 
\begin{equation}
b_{1}=\left(3\cdot I_{1}+4\cdot I_{2}+3\cdot I_{3}+R_{1}I_{1}+R_{2}I_{1}\right)/\label{eq:b1}
\end{equation}
\[
\left(\sum_{k=1}^{f}k\cdot I_{k}+R_{1}I_{1}+R_{2}I_{1}+2\cdot R_{1}I_{2}+2\cdot R_{2}I_{2}+2R_{2}I_{1}\right),
\]
\begin{equation}
b_{2}=\frac{2\cdot R_{2}+R_{2}I_{1}}{R_{2}+R_{2}I_{1}+2\cdot R_{2}R_{2}+R_{2}I_{2}+R_{2}R_{1}},\label{eq:b2}
\end{equation}
\begin{equation}
b_{3}=\frac{R_{2}^{2}}{R_{2}^{2}+2R_{2}I_{1}}.\label{eq:b3}
\end{equation}
The denominator in the three equations above contains all star connectivities
that can be connected with a given star by $n$ bonds between both
molecules, while the enumerator contains only that fraction of connectivities
that contains at least one unreacted group multiplied by the particular
number of unreacted groups.

The average distance between the star centers is reduced with increasing
number $n$ of connections between the stars. Assuming random walk
conformations and Gaussian elasticity of strands, loop formation is
additionally proportional to $[(1+1/n)/2]^{-3/2}$. Thus, we find
\begin{equation}
k_{2,n}(c)=2^{-3/2}b_{n}\left[\left(1+1/n\right)/2\right]^{-3/2}k_{1,f-2}(c),\label{eq:k2f1}
\end{equation}
which we define here for a single unreacted group at the selected
star and one connected star using $n$ bonds. Note that the proper
concentration dependence enters implicitly via the concentration dependence
of $k_{1,f-2}$.

For some star connectivities, the particular number of connected stars
with $n$ bonds and the numbers of unreacted groups $f-m$ is different
from one. This is then explicitly denoted as individual prefactor
of the rates $k_{2,n}$ at equations (\ref{eq:dr2}) to (\ref{eq:dr1r2})%
\footnote{For instance, the connectivity $I_{2}$ has two single connections
and two unreacted groups, which leads to an extra factor of $4$ for
$k_{2,1}$ at equation \ref{eq:di2}.%
}. 

With the above set of equations we can compute the ratios among all
rate constants as function of conversion, and thus, the problem can
always be solved by numerical integration. Note that above only the
ratio of $k_{1,f-2}(c^{*})/k_{0,f-1}(c^{*})$ (i.e. single pair intra-molecular
vs. single pair inter-molecular reaction rate) has to be fixed by
comparison with the simulation data. All other reaction rates are
computed as multiples of either $k_{1,f-2}$ or $k_{0,f-1}$, and
thus, this ratio is the \emph{only} adjustable parameter for all computations.

There is a large body of works that derive rate equations similar
to our discussion above as summarized, for instance, in the review
of Kuchanov \cite{Kuchanov}. The largest overlap exists with the
work of Dusek \cite{Dusek} and Sarmoria et al. \cite{Pereda,Sarmoria,Sarmoria2}.
Because of computing expectation values for the intra-molecular reactions
(equations \ref{eq:b1} - \ref{eq:b3}) our approach is formally equivalent
to model A of \cite{Sarmoria2}. The main differences to the present
work are a) the work of Sarmoria et al. does not consider conformational
changes of the polymers upon dilution, b) conformational changes due
to the formation of cycles are also neglected (cf. equation \ref{eq:k2f1}),
and c) the spanning tree approximation is problematic beyond the gel
point \cite{Sarmoria2}, while the approach of the present work investigates
rings of short length and thus, does not suffer from a diverging fraction
of intramolecular reactions. The limitation to short cycles is clearly
problematic close to the gel point, however, the comparison with the
simulation data will show that our approximation is sufficient for
well-developed networks at essentially all experimentally important
concentrations.

Note that for a discussion of samples close to the gel point, the
set of rate constants above can be readily extended to incorporate
cycles of larger size up to critical generation $i_{c}$ at which
there is the transition from mainly branching structures to a 3d-like
behaviour of the network structure \cite{lang2004promotion}. The
approach breaks down when modeling cyclic structures containing more
than $i\ge i_{c}$ molecules, since the contribution of intramolecular
reactions is diverging. We also have to clarify that our work represents
an enhanced mean-field approach that aims to correctly approximate
the effects of cycles in the networks, but does not consider fluctuations
of any kind. Therefore, it can only be applied as reasonable approximation
outside of the Ginzburg zone around the critical point \cite{Rubinstein}.

\section{Rate equations for star connectivities\label{sec:Rate-equations-for}}

For comparing simulation data and theory, we integrate here with respect
to conversion. Let $dp$ be an infinitesimal change in conversion
$p$. Using the above reaction rates we obtain for $f=4$ for the
entirely branching units $I_{i}$
\begin{equation}
\frac{dI_{0}}{dp}=-(k_{0,0}+k_{1,0})I_{0}\label{eq:di0}
\end{equation}
\begin{equation}
\frac{dI_{1}}{dp}=k_{0,0}I_{0}-(k_{0,1}+k_{1,1}+3\cdot k_{2,1})I_{1}\label{eq:di1}
\end{equation}
\begin{equation}
\frac{dI_{2}}{dp}=k_{0,1}I_{1}-(k_{0,2}+k_{1,2}+4\cdot k_{2,1})I_{2}\label{eq:di2}
\end{equation}
\begin{equation}
\frac{dI_{3}}{dp}=k_{0,2}I_{2}-(k_{0,3}+3\cdot k_{2,1})I_{3}\label{eq:di3}
\end{equation}
\begin{equation}
\frac{dI_{4}}{dp}=k_{0,3}I_{3}.\label{eq:di4}
\end{equation}
For the stars containing branching connections $I_{i}$ and loops
$R_{1}$ we get
\begin{equation}
\frac{dR_{1}}{dp}=k_{1,0}I_{0}-(k_{0,2}+k_{1,2})R_{1}\label{eq:dr1}
\end{equation}
\begin{equation}
\frac{dR_{1}I_{1}}{dp}=k_{1,1}I_{1}+k_{0,2}R_{1}-(k_{0,3}+k_{2,1})R_{1}I_{1}\label{eq:dr1i1}
\end{equation}
\begin{equation}
\frac{dR_{1}I_{2}}{dp}=k_{0,3}R_{1}I_{1}+k_{1,2}I_{2}\label{eq:dr1i2}
\end{equation}
\begin{equation}
\frac{dR_{1}^{2}}{dp}=k_{1,2}R_{1},\label{eq:d2r1}
\end{equation}
for the stars containing branching connections $I_{i}$ and loops
$R_{2}$
\begin{equation}
\frac{dR_{2}}{dp}=3\cdot k_{2,1}I_{1}-(k_{0,2}+k_{1,2}+k_{2,2})R_{2}\label{eq:dr2}
\end{equation}
\begin{equation}
\frac{dR_{2}I_{1}}{dt}=k_{0,2}R_{2}+4\cdot k_{2,1}I_{2}-(k_{0,3}+k_{2,1}+k_{2,2})R_{2}I_{1}\label{eq:dr2i}
\end{equation}
\begin{equation}
\frac{dR_{2}^{2}}{dp}=k_{2,2}R_{2}-(k_{0,3}+k_{2,3})R_{2}^{2}\label{eq:d2r2}
\end{equation}
\begin{equation}
\frac{dR_{2}I_{2}}{dp}=k_{0,3}R_{2}I_{1}+3\cdot k_{2,1}I_{3}\label{eq:dr2i2}
\end{equation}
\begin{equation}
\frac{dR_{2}R_{2}}{dp}=k_{2,1}R_{2}I_{1}\label{eq:dr2r2}
\end{equation}
\begin{equation}
\frac{dR_{2}^{2}I_{1}}{dp}=k_{0,3}R_{2}^{2}+k_{2,2}R_{2}I_{1}\label{eq:2dr2i}
\end{equation}
\begin{equation}
\frac{dR_{2}^{3}}{dp}=k_{2,3}R_{2}^{2},\label{eq:d3r2}
\end{equation}
and for the stars containing loops of both types $R_{1}$ and $R_{2}$
\begin{equation}
\frac{dR_{1}R_{2}}{dp}=k_{1,2}R_{2}+k_{2,1}R_{1}I_{1}.\label{eq:dr1r2}
\end{equation}

\begin{figure}
\includegraphics[angle=270,width=1\columnwidth]{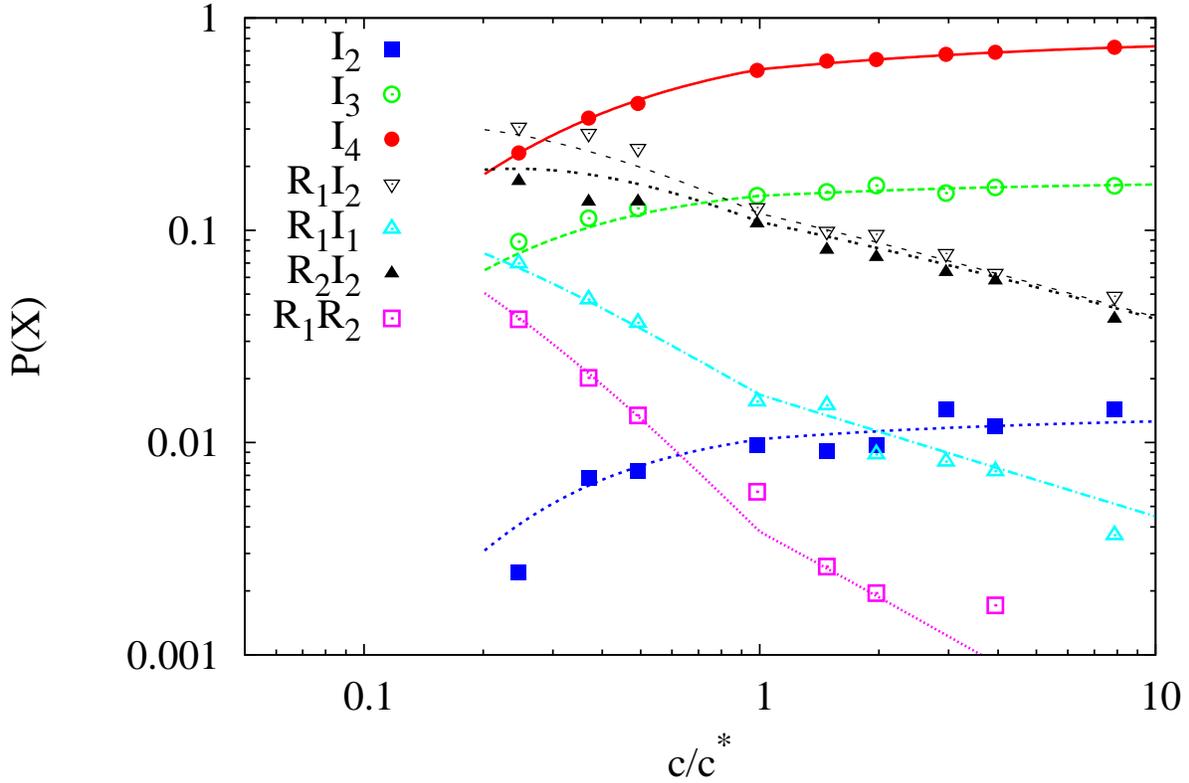}

\caption{\label{fig:species-A}Frequency of occurrence of frequent star connectivities
as function of concentration for A networks. Data points: simulation
data, lines are obtained by numerical integration of equations (\ref{eq:di0})
to (\ref{eq:dr1r2}) using $p=0.95$ and the additional factor of
$6/5$ for ring formation to match the data. The kink in the theoretical
lines is due to applying different regimes for star conformations.}
\end{figure}

Note that the above rate equations were written for A systems. AB
systems are readily modeled by putting all reaction rates $k_{1,i}=0$
and by computing only species that do not contain $R_{1}$. Then,
only reactions between species of the right half of the reaction diagram
are taken into account.

\begin{figure}
\includegraphics[angle=270,width=1\columnwidth]{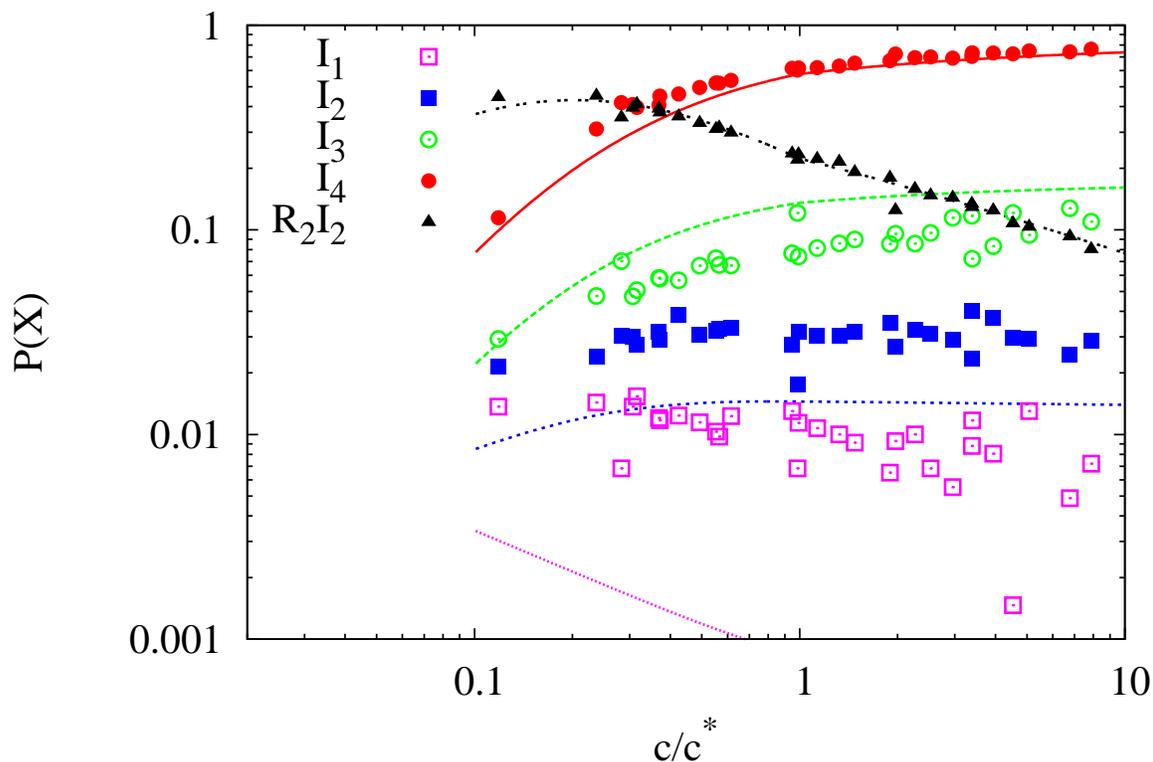}

\caption{\label{fig:species-AB-1}Frequency of occurrence of the most predominant
star connectivities as function of concentration for AB networks.
Simulation data (data points) and computation (lines) using the same
(fit-)parameters as in Figure \ref{fig:species-A} and considering
the smaller density of A and B groups.}
\end{figure}

\begin{figure}
\includegraphics[angle=270,width=1\columnwidth]{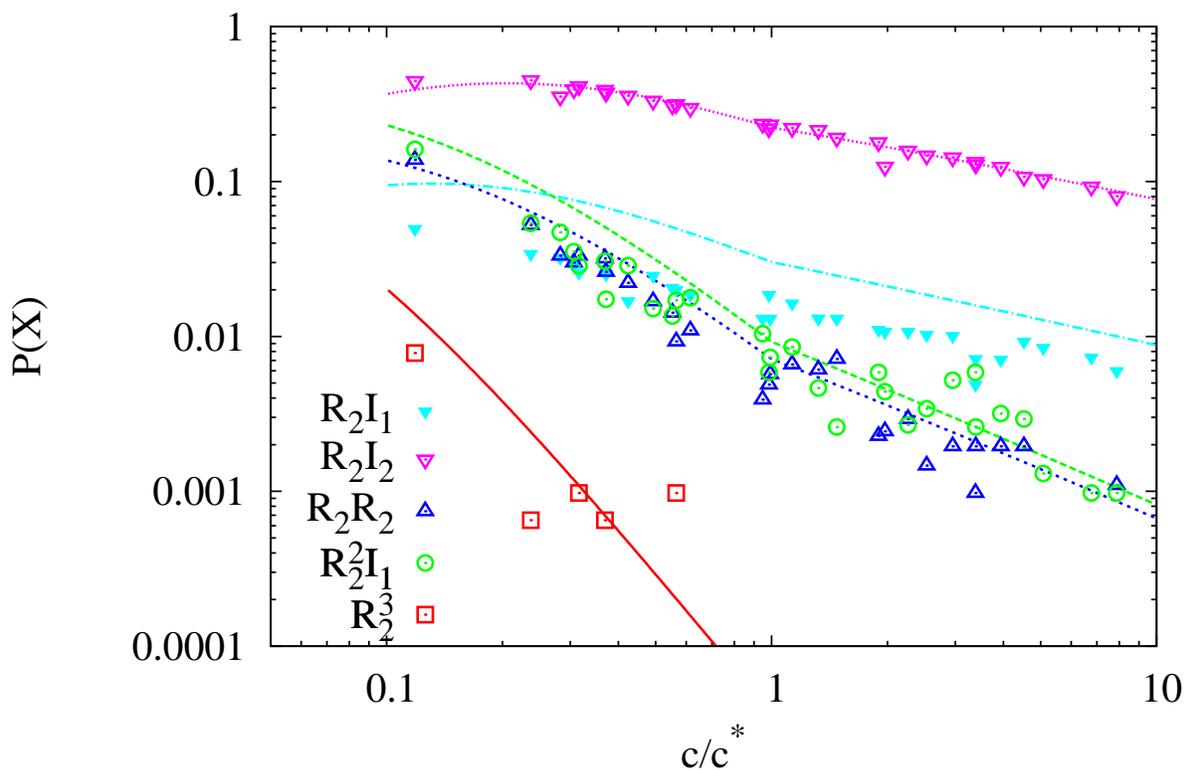}

\caption{\label{fig:Frequency-of-occurrence}Frequency of occurrence of star
connectivities $R_{2}I_{2}$, $R_{2}R_{2},$ $R_{2}^{2}I_{1},$ and
$R_{2}^{3}$ with four reacted arms in AB networks. The connectivity
$R_{2}I_{1}$ (full symbols) contains one unreacted arm and is shifted
with respect to the theoretical predictions similar to $I_{3}$ (see
Figure \ref{fig:species-AB-1}). }
\end{figure}

Figure \ref{fig:species-A} compares the results of the numerical
integration of the above rate equations with the data of the most
frequent structures in the A network series. For any star connectivity,
we find excellent agreement. Weak deviations are visible below the
overlap concentration, which might be caused by density fluctuations
of stars in space. For loops $R_{2}$ the theory slightly overestimates
loop formation. Here, the repulsion among stars and partial disinterpenetration
near $c^{*}$ \cite{Schwenke} was ignored in the rate equations and
the data were adjusted to fit overall formation of loops $R_{1}$.
Note that for figures \ref{fig:species-A}-\ref{fig:Frequency-of-occurrence}
data in the range of $P(X)=0.001$ refers to single events per sample.
Due to the limited number of about $4000$ stars per sample, the relative
error for the data is of order $[4000\cdot P(X)]^{-1/2}$. Therefore,
the data of Figure \ref{fig:species-A} suggest that the above mean-field
approximation is well suited to describe the distribution of the predominant
reactive species in $f$-functional homopolymerizations at concentrations
from $c^{*}$ up to melt conditions.

The same parameters for loop formation as for A networks can be chosen
to compare with AB co-polymer networks. Due to the stoichiometry of
the samples, the total density of reactive groups is twice the density
of reactive A or B groups, and thus, the formation of loops of same
size as in A networks is roughly doubled in AB networks. Since A reacts
exclusively with B, there are only rings containing an even number
of molecules possible. Figure \ref{fig:species-AB-1} compares our
predictions with the simulation data. Here, composition fluctuations
of the local volume fractions of A and B species freeze in upon cross-linking
and lead to regions with fluctuating conversion, depending on the
local stoichiometry of the species \cite{lang2003effect}. This is
also indicated at Figure \ref{fig:The-gyration-volumes}: within the
black circle, there is a larger number of red molecules, which leads
to a local imbalance of reactive groups. Interestingly, these fluctuations
leave the formation of loops $R_{2}I_{2}$ mainly unaffected. The
reason for this observation is that loops $R_{2}$ are formed between
A and B and thus (as all reacted groups) are part of the homogenously
distributed A and B groups, while the remaining unreacted dangling
arms at high conversion are mainly in volumes with non-stoichiometric
distribution of A and B groups. Therefore, the frequency of connectivities
that are incompletely reacted is mainly affected by local fluctuations
of A vs. B reactive groups. The deviations between simulation data
and theoretical prediction indicate that mean-field estimates for
co-polymerizations need to be improved to take into account the effect
of composition fluctuations of the different species. These fluctuations
depend strongly on mixing and total reaction rate \cite{lang2003effect}
and should show a clear dependence on the functionality of the molecules.

The species $R_{2}R_{2}$, $R_{2}^{3}$, and $R_{2}^{2}I_{1}$ have
an additional dependence on the number of existing bonds between a
pair of molecules and depend strongly on the estimate of the number
of un-reacted groups on connected molecules. Therefore, the data on
these structures is the best choice for testing equation (\ref{eq:k1m})
and (\ref{eq:k2f1}). Figure \ref{fig:Frequency-of-occurrence} shows
that our computations yields a rather good over all agreement for
all species that have four arms reacted. Species containing rings
with less than four reacted arms show quantitatively the same shift
in the frequency of occurrence as the species $I_{j}$ with the same
number of reacted arms $j<f$ in Figure \ref{fig:species-AB-1}. As
an example we included the most frequent species $R_{2}I_{1}$. Note
that species $R_{2}^{3}$ leads to an extra amount of sol that is
exclusively made of short cycles.

In contrast to previous work \cite{Rankin,Kurdikar,BenNaim,Sarmoria},
the simulations of our work explicitly model conformations and dynamics
of all monomers by taking into account excluded volume, entanglements
and the embedding of the monomers in space. Reactions occur on the
course of motion and collision of reactive groups in space. Therefore,
the data of our simulations are independent of the mean field assumptions
made for computing the rate equations and thus, serve as a critical
test of our approach. The good agreement between data and theory shows
that our approach is suitable to describe the short range connectivity
based on a nearest neighbour analysis. The above detailed analysis
of short loops in networks also reveals that for both A and AB networks
the loops $R_{1}$ and $R_{2}$ (if existing) are dominating for the
parameters of the present study. More complicated loop-structures
that may have a different concentration dependence, see Figure \ref{fig:species-A}
to \ref{fig:Frequency-of-occurrence} are clearly less pronounced
but become important for very low concentrations. Therefore, the simple
scaling approach in our previous work \cite{Schwenke} is justified
to approximate the total amount of loops in the samples for the current
set of parameters. We further note that the simulation results of
our proceeding work were in good qualitative agreement with recent
experimental data \cite{Lange2}. Therefore, we assume that the present
study is fundamental for a detailed analysis of the structure of these
four arm star tetra-PEG networks.

\section{Branching distributions and network properties\label{sec:Branching-distributions-and}}

Above we showed that our approach is suitable to describe the formation
of short loops in the networks. The elasticity of a polymer strand,
however, depends on the global embedding of this strand into the network
structure. Therefore, we test in the current section, whether our
mean-field description of short cyclic defects is sufficient to conclude
for the global structure of the networks that manifests itself, for
instance, in the amount of active material.

According to the definition of the connectivities in Figure \ref{Flo:Reaction diagram},
the mole fractions of stars $a_{i}$ which can have $i$ connections
to the incipient gel is given by
\begin{equation}
a_{0}=I_{0}+R_{1}^{2}+R_{2}^{3}+R_{1}\label{eq:B0}
\end{equation}
\begin{equation}
a_{1}=I_{1}+R_{1}R_{2}+R_{2}+R_{1}I_{1}+R_{2}^{2}\label{eq:B1}
\end{equation}
\begin{equation}
a_{2}=I_{2}+R_{1}I_{2}+R_{2}^{2}I_{1}+R_{2}R_{2}+R_{2}I_{1}\label{eq:B2}
\end{equation}
\begin{equation}
a_{3}=I_{3}+R_{2}I_{2}\label{eq:B3}
\end{equation}
\begin{equation}
a_{4}=I_{4}.\label{eq:B4}
\end{equation}

The distribution of mole fractions $a_{i}$ is the key quantity to
operate with the work of Miller \& Macosko \cite{miller1976new}.
In our approach, we consider a cyclic structure $R_{1}$ as a reduction
of the functionality by $2$ and loops $R_{2}$ as reductions of the
effective functionality by 1 as used in the above equations. For connectivity
$R_{2}^{3}$ we consider that these species can only be part of sol
for $f=4$.

Let $P(F_{A}^{out})$ denote the probability of finding a finite chain
starting at a randomly selected arm. Then we can write following the
ideas of Ref. \cite{miller1976new}
\begin{equation}
P(F_{A}^{out})=pP(F_{A}^{in})+1-p\label{eq:mm1}
\end{equation}
and 
\begin{equation}
P(F_{A}^{in})=\sum_{i}a_{i}P(F_{A}^{out})^{i-1}.\label{eq:mm2}
\end{equation}
Since $P(F_{A}^{out})=1$ is always a solution, the desired root%
\footnote{Note that the first term of the solution is missing at equation (24)
of Ref. \cite{miller1976new}.%
} to solve the above set of equations is 
\begin{equation}
P(F_{A}^{out})=-\frac{a_{3}}{2a_{4}}-\frac{1}{2}\label{eq:mm3}
\end{equation}
\[
+\left(a_{3}^{2}-a_{4}\left(3a_{4}+2a_{3}+4a_{2}-4/p'\right)\right)^{1/2}/2a_{4},
\]
provided that this solution is between 0 and 1. Since above we included
also incompletely reacted stars into the distribution $a_{i}$ and
we are only concerned with stoichiometric samples here, we can put
$p'=1$.

The probability that a unit of $0\le i\le f$ potential connections
has $k\le i$ bonds that are not finite is given by the binomial distribution
\begin{equation}
P(X_{k,i})\approx\left(\begin{array}{c}
i\\
k
\end{array}\right)P(F_{A}^{out})^{i-k}(1-P(F_{A}^{out}))^{k}.\label{eq:Xf}
\end{equation}
Note that the equation above is exact, if there are no correlations
among different types of connected stars. Since rings of type $R_{2}$
always introduce correlations among the functionalities of connected
stars, we implicitly test below by comparing with simulation data
whether ignoring these local correlations is a good approximation
for global connectivity, since the simulation results are subject
to these correlations.

In the classical definition of the active material by Scanlan \cite{scanlan1960effect}
and Case \cite{case1960branching}, the global connectivity of a given
branching point is analyzed: If there are at least three independent
paths to the network, the branching point is called active and any
chain (or series of chains) connecting two active branching points
is part of the elastically active material. Any part of the network
that is attached to the rest of the sample by only one connection
contributes to the elastically inactive ``dangling'' material. Thus,
for estimating the amount of active material, we have to first compute
the number fraction of stars that have exactly $k$ non-finite independent
connections to gel:
\begin{equation}
X_{k}=\sum_{i=k}^{f}A_{i}P(X_{k,i}).\label{eq:Xk}
\end{equation}
\begin{figure}
\includegraphics[angle=270,width=1\columnwidth]{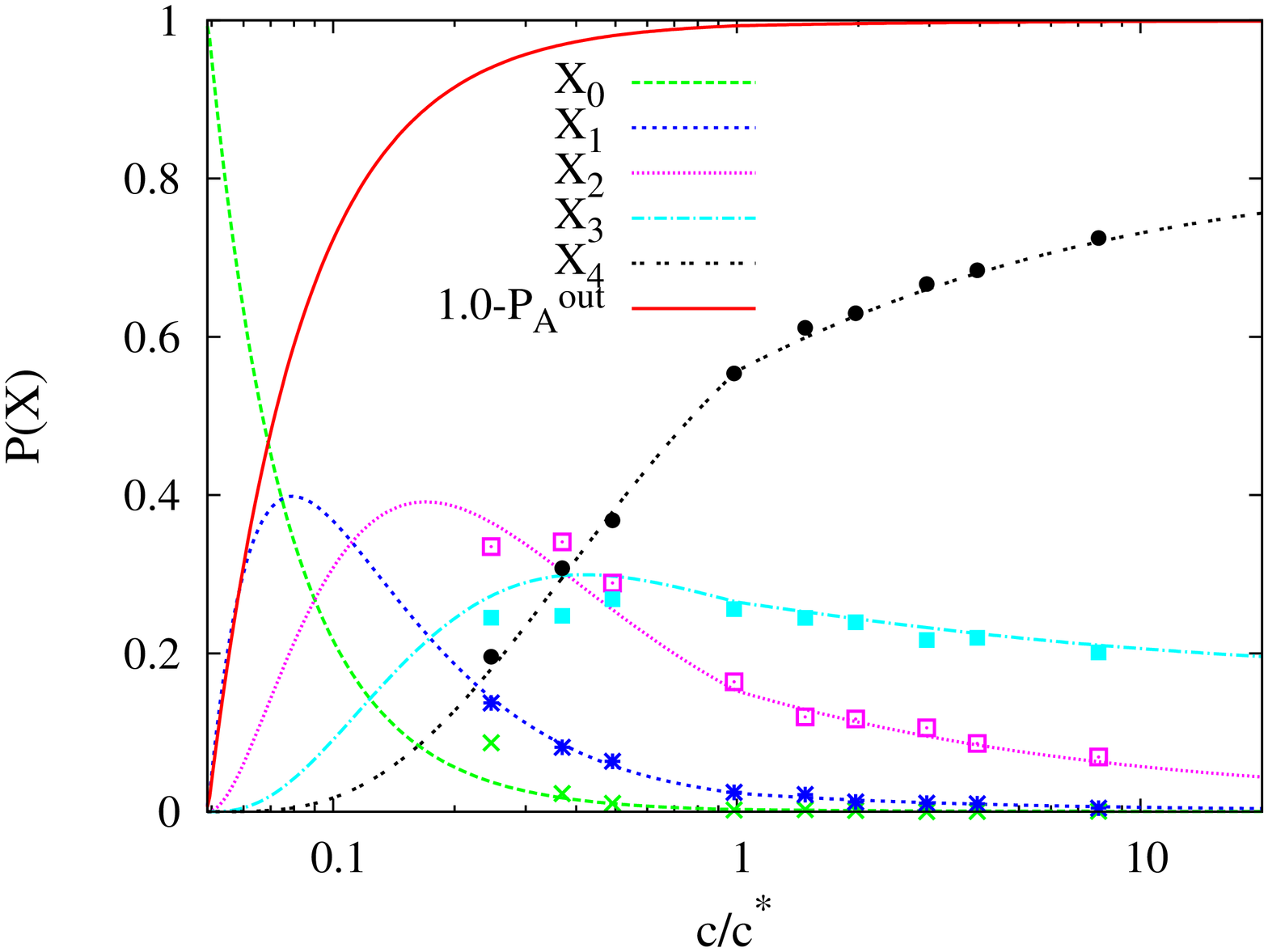}\caption{\label{fig:X_A}Number fractions of stars $X_{i}$ with $i$ independent
paths to the incipient gel for A networks of $f=4$ at $p=0.95$ (data
points) compared with the theoretical prediciton. }
\end{figure}

\begin{figure}
\includegraphics[angle=270,width=1\columnwidth]{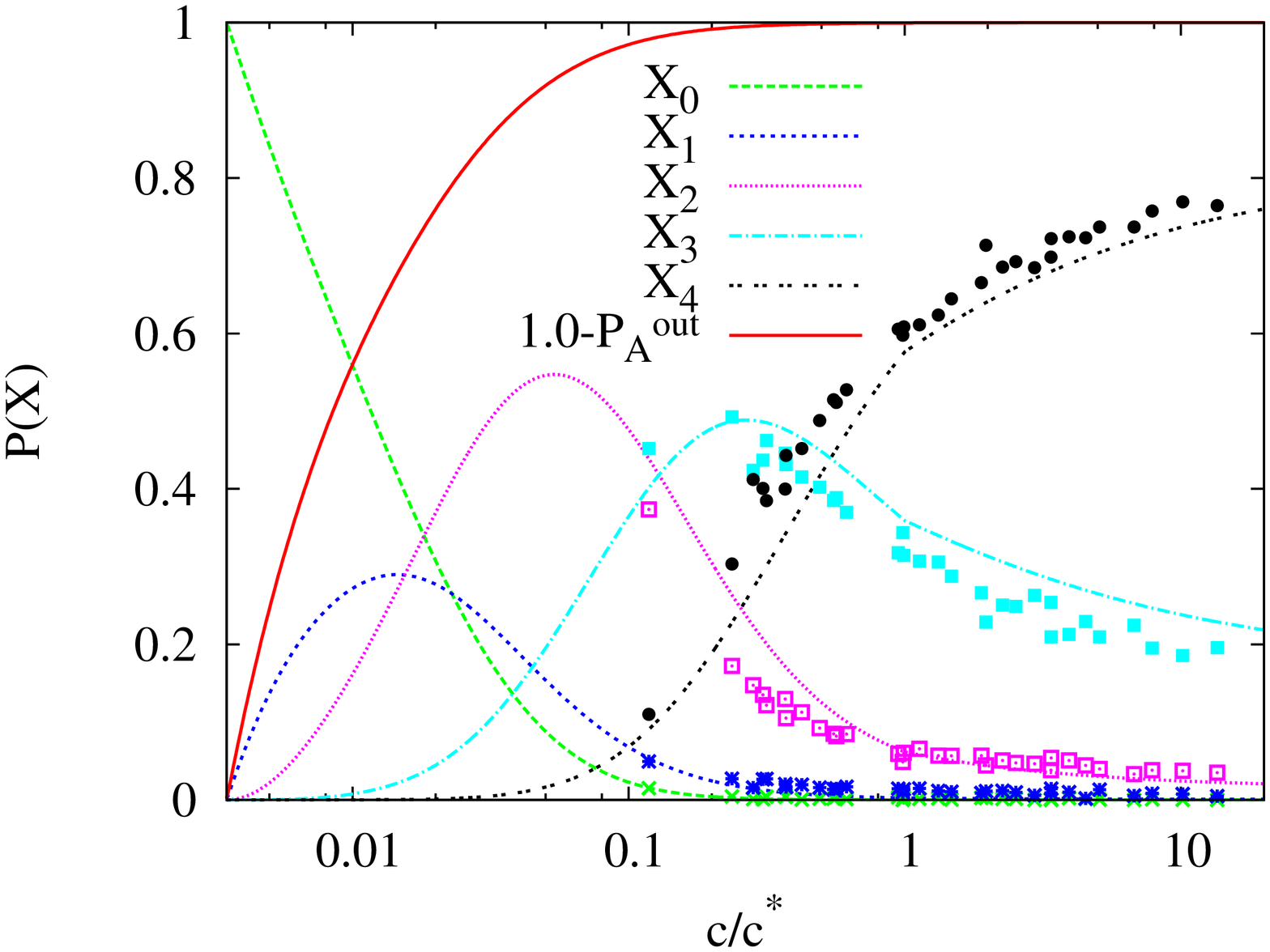}\caption{\label{fig:X_AB}Number fractions of stars $X_{i}$ with $i$ independent
paths to the incipient gel for AB networks of $f=4$ at $p=0.95$
(data points) compared with the theoretical prediction. }
\end{figure}

These number fractions are shown in Figure \ref{fig:X_A} and \ref{fig:X_AB}
for A and AB networks respectively. In our approach we consider only
$R_{1}$ and $R_{2}$ formation and neglect correlations among rings
and density fluctuations of stars in space. Therefore, the estimates
for the gel point (as suggested by the intersection of the probability
of finding an infinite chain starting at a randomly selected arm,
$(1-P_{A}^{out})$ and the $x$-axis) are lower boundaries for the
gel point and the distributions clearly below $c^{*}$ have to be
considered with care. Note that the kinks in the distribution function
are due to different approximations above and below the overlap concentration
(cf. equation (\ref{eq:k1m}) and (\ref{eq:k1m2})).

The data for the A networks in Figure \ref{fig:X_A} are in rather
good agreement with the theoretical predictions. Therefore, we conclude
that correlations among the functionalities of connected stars can
be ignored in first approximation. Concentration fluctuations of stars,
as for instance important for the increased formation of $X_{0}$
below $c^{*}$ are more important to be considered. The data of the
AB networks in Figure \ref{fig:X_AB} shows on the other hand, that
the changes in the local connectivity caused by concentration fluctuations
of A and B stars also lead to a systematically modified global connectivity.
Additionally, using an AB type reaction leads to a large shift of
the predicted minimum concentration for gel formation at $p=0.95$
as compared to A networks. This is mainly caused by the absence of
loops $R_{1}$.

Next, we analyze the weight fractions of sol, the active, and the
dangling material. The weight fraction $w_{a}$ of active material
among all polymer is the weight fraction of effective strands
\begin{equation}
w_{e}=\sum_{i=2}^{f}iX_{i}/f\label{eq:nue_e}
\end{equation}
plus one additional strand per loop $R_{2}$ inside the active material.
Thus, we have to consider the remaining connections attached to the
loop(s) between the two stars in order to estimate whether this loop
is active or not. The additional active material stored in loops $R_{2}$
is, therefore, given by 
\begin{equation}
w_{R_{2}}\approx P(X_{2,2})R_{2}I_{1}/4+P(X_{3,3})R_{2}I_{2}/4\label{eq:wr2}
\end{equation}
\[
+P(X_{2,3})R_{2}I_{2}/6+P(X_{2,2})2R_{2}I_{1}/2+P(X_{2,2})R_{2}R_{2})/2.
\]
The total amount of the active material is then 

\begin{equation}
w_{act}=w_{e}+w_{R_{2}}.\label{eq:wa}
\end{equation}
Note that $2w_{R_{2}}/w_{act}$ is, therefore, the weight fraction
of loops $R_{2}$ inside the active material.

According to our definitions,
\begin{equation}
w_{sol}=A_{0}\label{eq:wsol}
\end{equation}
and therefore, the weight fraction of dangling material can be computed
as 
\begin{equation}
w_{dangl}=1-w_{act}-w_{sol}.\label{eq:wdangl}
\end{equation}

\begin{figure}
\includegraphics[angle=270,width=1\columnwidth]{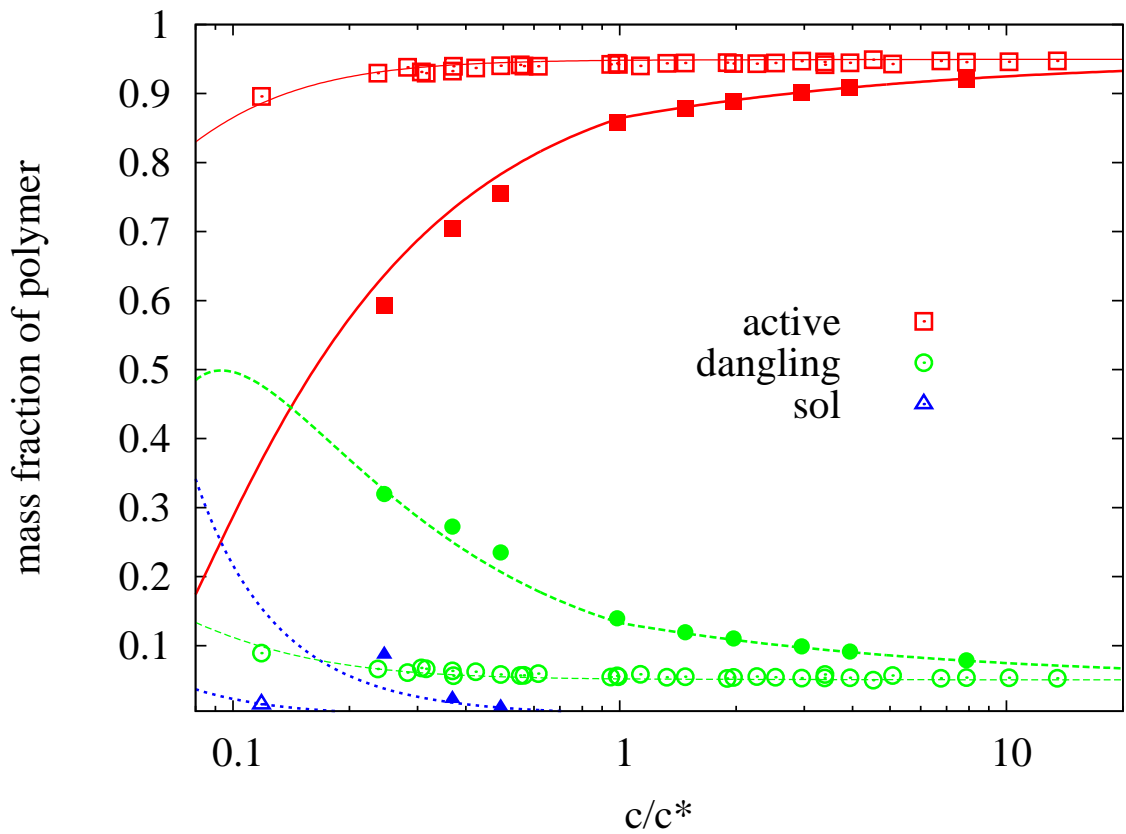}

\caption{\label{fig:Polymer-mass-fractions-1-1-1}Polymer mass fractions of
dangling material (circles), sol (triangles), and active material
(squares) of A (full symbols) and AB networks (hollow symbols) as
function of concentration. Thick lines represent estimates for A networks,
thin lines for AB networks.}
\end{figure}

The amount of active and dangling material as well as sol is shown
in Figure \ref{fig:Polymer-mass-fractions-1-1-1} for A and AB networks.
The striking result of this analysis is that even at concentrations
of $c^{*}/10$ nearly all polymer material is active for the AB networks
in large contrast to A networks. This stems from the fact that that
loops $R_{1}$ are missing in AB networks and the predominant non-ideality
of network structure is the formation of loops $R_{2}$ which effectively
leads to a reduction of average functionality (here from $4$ to $3$)
but otherwise no impact on the amount of the active material. Note
that this observation requires $f\ge4$. Over all, Figure \ref{fig:Polymer-mass-fractions-1-1-1}
suggests that the network stability is nearly constant over a wide
range of $c$, if the same conversion $p$ was achieved among all
samples. Therefore, this result is in striking agreement with the
nearly constant breaking strain as observed in the experiments of
Sakai \cite{Sakai08}. However, multiple links lead to a reduction
of the average effective functionality of the stars and thus, to a
modification of phantom modulus different to changes in the amount
of the active material as discussed in the following section.

\section{Application to Network Elasticity}

Flory already argued \cite{flory1976statistical}, that finite cycles
would diminish the elastic response of the network. This effect is
most pronounced for the shortest non-dangling rings and ignorable
for larger ring sizes, since the computation of the cross-link fluctuations
runs over a geometric series, which quickly converges against its
asymptote. In fact, the derivation of the phantom modulus 
\begin{equation}
G_{ph,id}=\frac{f-2}{f}\cdot\nu k_{B}T=\frac{f-2}{f}\cdot\frac{k_{B}T}{\mbox{v}_{mon}N}\label{eq:phantom und so-1}
\end{equation}
for ideal $f$-functional networks without defects requires the assumption
that all cycles are closed at infinity \cite{flory1976statistical}.
Here, $\nu$ is the number density of network strands, $k_{B}$ the
Boltzmann constant, $T$ temperature, and $\mbox{v}_{mon}$ the volume
of a monomer.

As compared to this ideal prediction, short loops inside the active
network structure will reduce the phantom modulus, even though these
rings are still actively deformed. In the discussion of Flory \cite{flory1976statistical},
a correction for these short circuits was omitted due to the large
overlap number, $P\gg1$, of active strands in networks crosslinked
at $c\gg c^{*}$, which reduces their frequency of occurrence. However,
this argument is questionable for polymer gels close to overlap concentration
$c^{*}$, for which only a small number of neighboring molecules exist.

In the following, we estimate the phantom modulus by using the number
average strand length $N_{av}$ and weight average active junction
functionality $\left\langle f_{a}\right\rangle $. Thereto, we interpret
multiple links as single connection of reduced degree of polymerization
between the same two star centers: a double link $R_{2}$ effectively
deforms as a single chain of $N/2$ monomers, while a triple link
$R_{2}^{2}$ deforms like a chain of $N/3$ monomers. A linear series
of two double links behaves like a single chain of $N$ monomers,
and so on. Since the effect of $R_{2}^{2}$ and a series of two double
links is rather small and almost mutually cancelling for most samples
of our study, we approximate the effect of multiple links onto average
strand lengths by replacing a fraction of $w_{R_{2}}/w_{act}$ active
connections with chains of $N/2$ monomers. This leads to an average
strand length of roughly $[1-w_{R_{2}}/2w_{act}]N$ monomers, whereby
$w_{act}$ is the weight fraction of active strands and $w_{R_{2}}$
half the weight fraction of the loops $R_{2}$ inside the active material
(cf. section \ref{sec:Branching-distributions-and} for the computation
of $w_{act}$ and $w_{R_{2}}$). 2-functional junctions as caused
by dangling loops for $f=4$ on the other hand effectively dilute
the concentration of branching points inside the active material by
a factor 
\begin{equation}
\alpha=\sum_{k=2}^{i}X_{k}/\sum_{k=3}^{i}X_{k}.\label{eq:alpha-1}
\end{equation}
Here, $X_{k}$ is the number fraction of units with $k$ non-finite
connections to the gel. The average active network strand is therefore
\begin{equation}
N_{av}\approx\alpha[1-w_{R_{2}}/2w_{act}]N.\label{eq:Nav-1}
\end{equation}
and the average active functionality for $f=4$ is given by 
\begin{equation}
\left\langle f_{a}\right\rangle =(3X_{3}+4X_{4})/(X_{3}+X_{4}).\label{eq:fa-1}
\end{equation}
Both expressions for $N_{av}$ and $\left\langle f_{a}\right\rangle $
then replace $N$ and $f$ at equation (\ref{eq:phantom und so-1}).

To decide whether $\left\langle f_{a}\right\rangle $ of $N_{av}$
is playing the leading role for the effect on modulus depends to a
large extent on whether it is an A or an AB network, in particular
for small $f=3,4$. For our series of data, the modification of $N_{av}$
and the reduction in the concentration of active junctions are responsible
for the reduction of the modulus of A networks above $c^{*}$, while
below $c^{*}$ the average functionality of active junctions is increasingly
affected for A and AB networks.

\begin{figure}[htbp]
\centering \includegraphics[angle=270,width=1\columnwidth]{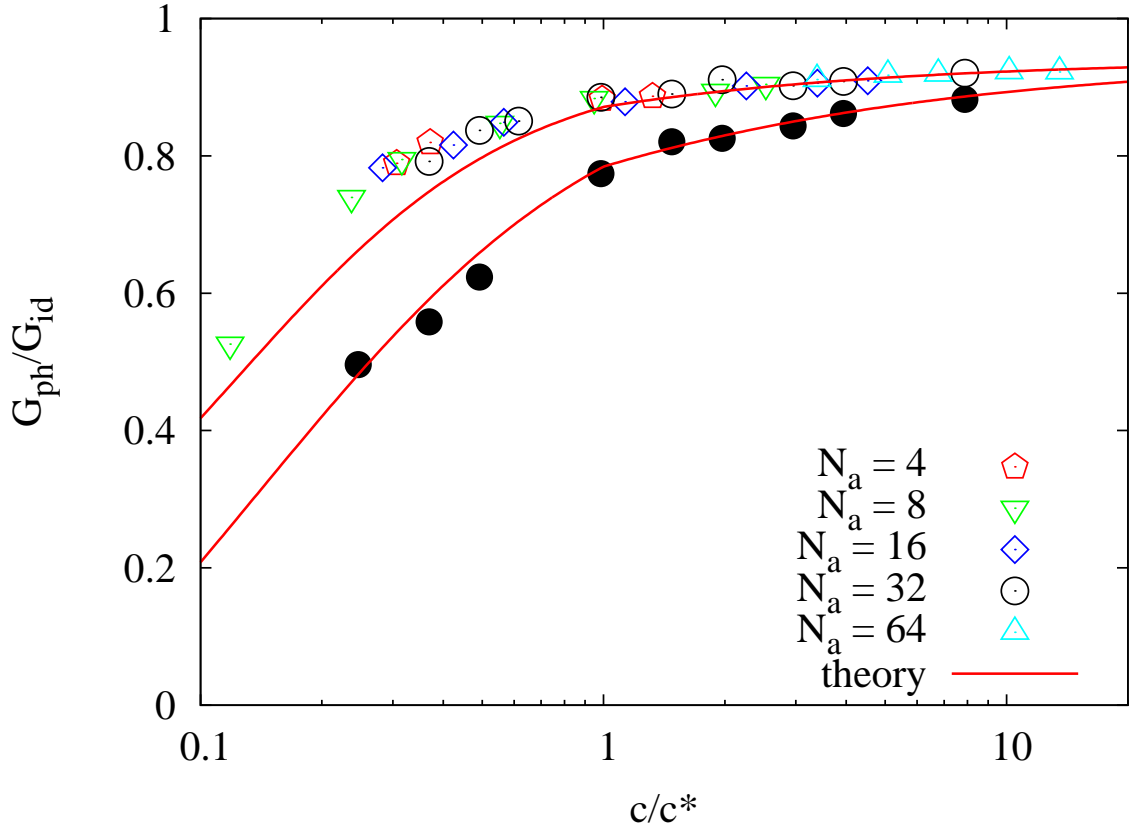}
\caption{The reduction of phantom modulus after consideration of rings $R_{2}$
and $R_{1}$ as function of the overlap number of stars $c/c*$. Hollow
symbols show the data of AB networks, full symbols data of A networks.
Data is normalized by the ideal phantom modulus $G_{id}$ of a 4-functional
network at same density.}

\label{fig:elactmatMM-1} 
\end{figure}

$G_{ph}$ was computed independently for the network structures of
the simulations and for the solutions of the rate equations of section
\ref{sec:Rate-equations-for}. The result of this computation as shown
at Figure \ref{fig:elactmatMM-1} indicates that the above estimate
agrees well with the data on the homopolymer networks and thus, correlations
among multiply connected molecules can be neglected for these samples.
The effect of fluctuations is still clearly visible for the AB networks.
Furthermore, we observe a clear drop of the phantom modulus of AB
networks as compared to ideal network structures starting at $c^{*}$
to lower concentrations. This drop sets in at much higher conversions
as the drop of the amount of active material, since it is mainly caused
by the reduction of the average functionality of the molecules.

\section{Summary}

We have presented a rate theory approach to describe network formation
of $f$-functional stars including the effect of short loops. We showed
that the only parameter for all computations is the ratio of $k_{1,f-2}(c^{*})/k_{0,f-1}(c^{*})$
(i.e. single pair intra-molecular vs. single pair inter-molecular
reaction rate), since all other reaction rates are functions of one
of these two reaction rates. This ratio is well approximated by the
ratio of the gyration volume and the accessible volume of star arms
$R_{g}^{3}/R_{e}^{3}$ up to a numerical fit-factor of 6/5 for the
data of our study. The concentration dependence of the reaction rates
is in rather good agreement with the scaling of the conformations
of the molecules. The observed small change of the scaling of loop
formation as function of concentration was attributed to a partial
disinterpenetration of the molecules, which has not yet been considered
by theory. Agreement between data and theory, thus, justifies the
neglect of diffusion for the discussion of network formation for networks
of four-arm stars. This observation is in agreement with a recent
experimental study on reaction kinetics\cite{Nishi}, which observed
no effects of steric hindrance or the gelation threshold, which both
are related to diffusion.

For the samples of our study, we could not find significant differences
as function of reaction rate. Therefore, it might be possible to ignore
the effect of dynamics on the structure of networks of star molecules,
if these were obtained from homogeneously mixed samples. The comparison
of the theoretical predictions with simulation data on homopolymer
A networks reveals that correlations among molecules can be ignored.
Also, for concentrations $c>c^{*}/4$ the effect of local fluctuations
of polymer density can be ignored and mean-field descriptions can
be rather precise for computing the amount of short cyclic structures
inside the networks, as we can judge from the data of the present
work.

Co-polymer networks made by exclusively reacting A with B species
on the other hand show a clear effect of composition fluctuations
that leads to fluctuations of the reactive groups of one species in
space, even though the total density of polymers might be equilibrated.
Mean-field descriptions of these samples are therefore less accurate.
Furthermore, experimental data is expected to be very sensitive with
respect of mixing both types of molecules. The data of the present
study does not allow to look with more detail into this problem, since
the main parameters for this effect are a) a amplitude of the initial
composition fluctuation, b) a variation of the over all reaction rate,
c) the functionality of the molecules. But our results suggest that
a systematic study will be essential for understanding the differences
between homo-polymer and co-polymer networks. The results for the
total amount of active material suggest a much more stable network
structure for very low concentrations in the case of AB-networks as
compared to homopolymer networks at same conversion. Apparently, this
effect arises from the missing $R_{1}$ loops in AB networks. But
the modulus of the networks is expected to show a stronger dependence
on decreasing concentration at $c<c^{*}$ in case of AB networks,
since there, first the functionality of the molecules is reduced before
the amount of active material is lowered. 

Finally, the striking differences between A and AB networks (and between
these samples and ideal networks) exemplify the importance of understanding
of network formation: even though the building blocks of the network
are of identical structure, the condition of linking exclusively A
with B groups implies clearly different network structures and thus,
material properties.

\section{Acknowledgement}

The authors thank the ZIH Dresden for a generous grant of computing
time. ML thanks the DFG for funding project LA2735/2-1.

\bibliographystyle{plain}
\bibliography{Literatur}

\section*{\newpage{}}
\section*{Table of Contents Graphic}
\emph{Reaction Diagram}

Michael Lang

Konrad Schwenke

Jens-Uwe Sommer

\begin{figure}[htbp]
\includegraphics[angle=0,width=\columnwidth]{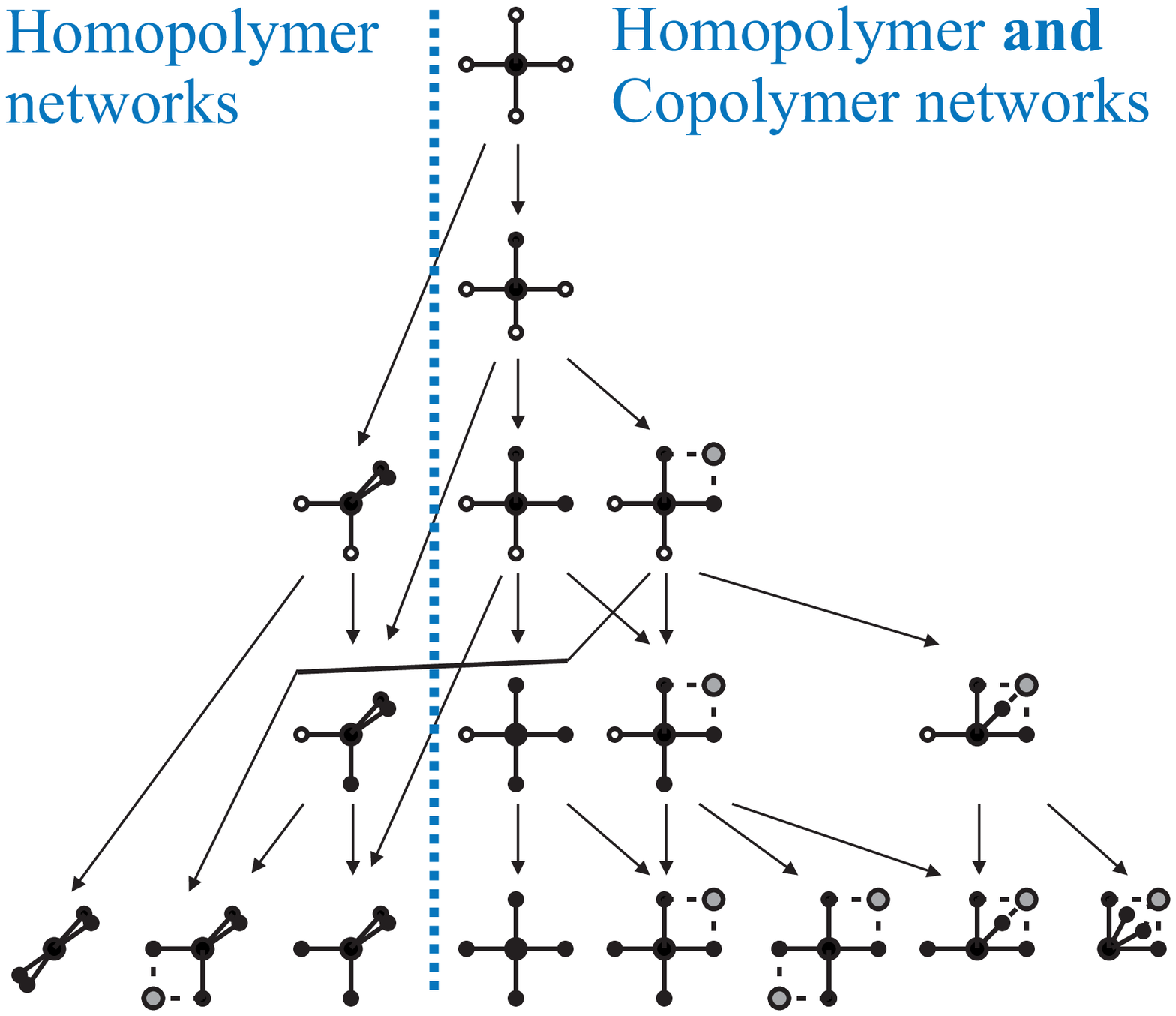}
\end{figure}
\end{document}